# A dynamical modeling to study the adaptive immune system and the influence of antibodies in the immune memory*

Alexandre de Castro[1,2**], Carlos Frederico Fronza[2] and Domingos Alves[2,3]

**Abstract.** Immunological systems have been an abundant inspiration to contemporary computer scientists. Problem solving strategies, stemming from known immune system phenomena, have been successfully applied to challenging problems of modern computing (MONROY, SAAB, GODÍNEZ, 2004). Simulation systems and mathematical modeling are also beginning use to answer more complex immunological questions as immune memory process and duration of vaccines, where the regulation mechanisms are not still known sufficiently (LUNDEGAARD, LUND, KESMIR, BRUNAK, NIELSEN, 2007).In this article we studied *in machina* a approach to simulate the process of antigenic mutation and its implications for the process of memory. Our results have suggested that the durability of the immune memory is affected by the process of antigenic mutation and by populations of soluble antibodies in the blood. The results also strongly suggest that the decrease of the production of antibodies favors the global maintenance of immune memory.

**Keywords:** immune memory, antibodies, B-cells,

## 1. INTRODUCTION

Natural computing brings together nature and computing to develop new computational tools for problem solving (CASTRO,2006). Approaches as evolutionary computing, neurocomputing and immunocomputing have found numerous applications in a variety of growing fields including engineering, computer science, and biological modeling (LIAN, LIOU, 2005; YOSHIMURA, 2006; AYMERICH, SERRA, 2006; LACERDA, SILVA, 2006; OISHI, YOSHIMURA, 2007; SINGH, MANI. GANGULI, 2007; YANG, TANG, HATSUKAMI, ZHENG, WOODARD, 2007; OISHI, YOSHIMURA, 2008; KERH, LAI, GUNARATNAM, SAUNDERS, 2008*).*

[1]Centro Nacional de Pesquisa Tecnológica em Informática para a Agricultura, Empresa Brasileira de Pesquisa Agropecuária (EMBRAPA), Campinas 13083-886, Brazil.

[2]Departamento de Informática em Saúde, Universidade Federal de São Paulo (UNIFESP), São Paulo 04023-062, Brazil

[3]Faculdade de Medicina de Ribeirão Preto, Universidade de São Paulo (USP), Ribeirão Preto 14049-900, Brazil.

**Corresponding author:

alexandre.castro@embrapa.br

In this scenario, immunocomputing may be an important tool to better to better understand the mammals' defense system and it can provide results difficult to be observed *in vivo.*

As a brief introduction to the immune system, we consider the immune responses mediated mainly by the lymphocytes B and T – responsible by the specific recognition of the antigens (strange molecules to the organism capable of being recognized by the immune system) – and by soluble molecules that this B lymphocytes secrete, the antibodies (ROITT, BROSTOFF, MALE, 1998).

In general, the immune system must present "virgin", "immune" and "tolerant" states and may present limits of memorization. In the "virgin" state the populations are all at a very low level, which means, with values of the order of the quantities randomly produced by the bone marrow. In the "immune"' state the population of B cells that specifically recognizes a kind of antigen remains at a determined level, even after the suppression of this antigen. In the "tolerant" state the populations of antibodies and of B cells did not respond to the presence of antigens or self-antigens (proteins from the individual's organism itself). However, there can be a positive selection of B cells naturally self-reactive,
indicating the existence of a subgroup of B lymphocytes subject to self reactivity (ROITT, BROSTOFF, MALE, 1998).

Among millions of kinds of B lymphocytes of the organism, each one with its specific antibody held on the membrane, only those which recognize a speciic antigen are stimulated. When this occurs, the B lymphocyte multiplies, originating a lineage of cells (clones) able to produce specific antibodies against the antigen that induced its multiplication.

The antibodies produced by a mature B lymphocyte – known as plasmacyte – are released in great amount in the blood. The multiplication continues as long as there are antigens able to activate them. As a determined kind of antigen is being eliminated from the body, the number of lymphocytes specialized in battle it also diminishes. However, a small population of these lymphocytes remains in the organism for the rest of the individual's life, constituting what is denominated immune memory. During the evolution of the immune system, an organism finds a given antigen repeatedly. The efficiency of the adaptive response to secondary encounters could be considerably increased by the storing of populations of cells that produce antibodies with high affinity to that antigen, denominated memory cells. Instead of "starting all over' each time a given antigenic stimulus is presented this strategy guarantees that the speed and efficiency of the immune response is enhanced after each infection (ROITT, BROSTOFF, MALE, 1998).

To better understand this process, two fundamental theories for the immune memory were developed. The first one considers that, after the expansion of the B cells, the formation of plasma cells and memory cells occur. According to F. M. Burnet (BURNET,1959), these memory cells would be remaining cells from an immunologic response that, supposedly, survive until the end of the individual's life – therefore with a life longer than the other cell of the organism.

The second theory, due to N. K. Jerne (JERNE, 1974), considers that the immune system presents memory and capacity of response for a second invasion of the same antigen, with a self-organization of the system, allowing the formation of cellular populations that last for a long time. In other words, this author theorizes that the populations survive, and not only a specific kind of cell with a lifespan longer than the one from other cells of the organism.

## 2. MATERIALS AND METHODS

In this paper, a computational model is presented to simulate the behavior of the immune system, considering structural mechanisms of regulation that were not included in the simplified model proposed by Lagreca (LAGRECA, ALMEIDA, SANTOS, 2001). In our approach we considered not only the antibodies linked to the surface of the B cells (surface receptors), but also the populations of antibodies soluble in the blood (antibodies secreted by mature B cells), making this model closer to the real behavior of the immune system (CASTRO, 2005; CASTRO, 2006; CASTRO, 2007; CASTRO, FRONZA, ALVES, 2009; CASTRO, FRONZA, GIACHETTO, ALVES, 2009).

Besides the differentiation of the B cells, the model exposed here allows representing the generation, maintenance and regulation of the immune memory in a more complete way, through a memory network, that combines the characteristics of Burnet's clonal selection theory and Jerne's network hypothesis, considering only idiotypic–antiidiotypic interactions (BURNET, 1959; JERNE, 1974).

In the model discussed here, the molecular receptors of the B cells are represented through bit-strings with diversity of $2^B$, where B is the number of bits in the string (CASTRO, 2005; CASTRO, 2006; CASTRO, 2007; CASTRO, FRONZA, ALVES, 2009; CASTRO, FRONZA, GIACHETTO, ALVES, 2009; PERELSON, WEISBUCH, 1997). The individual components of the immune system represented in the model are the B cells, the antibodies and the antigens. The B cells (clones) are characterized by their surface receptor and modeled by a binary string.

The epitopes – portions of an antigen that can be linked by the B cell receptors (BCR) – are also represented by bit strings. The antibodies have receptors (paratopes) that are represented by the same bit-string that models the BCR of the B cell which produced them.

Each string (shape) is associated to an integer σ ( $0 \leq \sigma \leq M = 2^B - 1$ ) which represents each on of the clones, antigens or antibodies. The neighbors to a given σ are expressed by the Boolean function $\sigma_i = (2^i \ 1 \text{xor} \sigma)$. The complementary form of σ is obtained by $\overline{\sigma} = M - \sigma$ and the time evolution of the concentrations of the several populations is obtained as a function of the integer variants σ and *t*, through direct iteration.

The equations that describe the behavior of the clonal populations are calculated through an iterative process, for different parameters and initial conditions:

$$m + (1-d) y(\sigma\alpha t) + b \frac{y(\sigma\alpha t)}{y_{tot}(t)} \zeta_{a_h}(\bar{\sigma}, t),$$

(1)

with the complementary shapes included in the term $\zeta_{a_h}(\bar{\sigma}, t)$,

$$\zeta_{a_h}(\bar{\sigma}, t) = (1-a_h)[y(\bar{\sigma}, t) + y_F(\bar{\sigma}, t) + y_A(\bar{\sigma}, t)] + a_h \sum_{i=1}^{B}[y(\bar{\sigma_i}, t) + y_F(\bar{\sigma_i}, t) + y_A(\bar{\sigma_i}, t)].$$

In these equations, $y_A(\sigma\alpha t)$ and $y_F(\sigma\alpha t)$ are, respectively, the populations of antibodies and antigens; $b$ is the proliferation rate of the B cells; $\bar{\sigma}$ and $\bar{\sigma_i}$ are the complementary shapes of $\sigma$ and of the nearest $B$ neighbors in the hypercube (with the *i*th bit flipped). The first term *(m)*, inside the curled brackets in equation (1), represents the production of the cells by the bone marrow and it is a stochastic variable. This term is small, but non-zero. The second term inside the curled brackets describes the populations that have survived to natural death *(d)*, and the third term represents the clonal proliferation due to iteration with complementary shapes (other clones, antigens or antibodies). The parameter $a_h$ is the relative connectivity among a determined bit-string and the neighborhood of its mirror image or complementary shape. When $a_h = 0.0$, only perfect complementary shapes are allowed. When $a_h = 0.5$, a string can equally recognize its mirror image and its first neighbors.

The factor $y_{tot}(t)$ is expressed by:

$$y_{tot}(t) = \sum_{\sigma}[y(\sigma, t) + y_F(\sigma, t) + y_A(\sigma, t)]$$

(2)

The time evolution of the antigens is determined by:

$$y_F(\sigma\alpha t+1) = y_F(\sigma\alpha t) - k \frac{y_F(\sigma\alpha t)}{y_{tot}(t)} \times$$
$$(1-a_h)[y(\bar{\sigma}, t) + y_A(\bar{\sigma}, t)] + a_h \sum_{i=1}^{B}[y(\bar{\sigma_i}, t) + y_A(\bar{\sigma_i}, t)],$$

(3)

where $k$ is the speed in which the populations of antigens or antibodies decrease to zero, which means, the antigen removal rate due to iterations with the populations of B cells and antibodies.

The populations of antibodies is described by a group of variable $2^B$ defined in a B-dimensional hypercube, interacting with the antigenic populations:

$$y_A(\sigma, t+1) = y_A(\sigma, t) + b_A \frac{y(\sigma, t)}{y_{tot}(t)} \times$$
$$(1-a_h) y_F(\bar{\sigma}, t) + a_h \sum_{i=1}^{B} y_F(\bar{\sigma_i}, t) - k \frac{y_A(\sigma, t)}{y_{tot}(t)} \zeta_{a_h}(\bar{\sigma}, t)$$

(4)

where the contribution of the complementary shapes $\zeta_{a_h}(\bar{\sigma}, t)$ is again included in the last term, $b_A$ is the antibody proliferation, and $k$ is the antibody removal rate, which measures its iterations with the other populations.

The populations of antibodies $y_A(\sigma\alpha t)$ (that represent the total number of antibodies) depend on the inoculated dosage of antigens. The factors $\frac{y_F(\sigma\alpha t)}{y_{tot}(t)}$ and $\frac{y_A(\sigma\alpha t)}{y_{tot}(t)}$ are the responsible by the control and decrease of the populations of antigens and antibodies, while the factor $\frac{y(\sigma(t)}{y_{tot}(t)}$ is the corresponding factor for the accumulation of the clone populations in the formation of the immune memory. The clonal population $y(\sigma(t)$ (normalized total number of clones) can vary since the value produced by the bone marrow *(m)* until its maximum value (in our model, he unity), since the Verhust factor limits its growth.

The Verhust factor produces a local control of the populations of clones (B cells), considering the several regulation mechanisms (CASTRO, 2009). However, the populations of B cells are strongly affected by the populations of antibodies soluble in the blood. This is the reason that leads us to include the term $\overline{\zeta_{a_h}}(\sigma,t)$ as an extra contribution in the set of maps previously coupled proposed by Lagreca (LAGRECA, ALMEIDA, SANTOS, 2001).

In order to properly study the time evolution of the components of the immune system, we define clone as being only a set of B cells. So, the population of antibody is treated separately in the present model.

The equations (1) to (4) form a set of coupled maps that describes the main interactions of the immune system among entities that interact through connections "key-lock" type, which means, entities that recognize each other specifically. This set of equations is solved iteratively, considering different initial conditions.

## 3. RESULTS

The simulations performed in this paper show the generation, maintenance and the regulation mechanisms of the immune memory, and the cellular differentiation, through idiotypic–antiidiotypic interactions, that combine the characteristics of the clonal selection theory and the immune network theory.

To show the extension of the validity of the model, the results of some simulations are presented. Immunization experiments, in which the several antigens, with fixed concentrations are injected in the body in an interval of 1000 time steps, in order to stimulate immune response. When a new antigen is introduced, its interaction with all the other components in the system is obtained through a random number generator.

The length of the B bit string was fixed in 12, corresponding to a potential repertory of 4096 cells and distinctive receptors. Injections of different antigens, in time intervals, corresponding to one period of life or to the entire life of the individual were given.

In the simulations, the value $d = 0.99$ was considered for the rate of natural death of the cells (apoptosis), and the proliferation rate of clones and antibodies, as being 2 and 100, respectively. For the connectivity parameter $a_h$ the value 0,01 was chosen; and the antibodies and antigens removal rate $(k)$ was fixed in 0,1, so that in each interval of 1000 time steps, the populations of antigens and antibodies disappear before the next antigen be inoculated.

In each inoculation, the same seed for the random number generator was used, so the different antigens are inoculated at the same order in all the simulations. Many simulations were performed, with antigen doses varying between 0.0001 and 1.5.

Next, the results of some simulations will be shown, with special emphasis to two intermediate values for the doses – 0.08 and 0.10 – in the region of coverage of the simulations. Although very close, these values present distinct results for the immune memory and, consequently, duration of vaccines. Little alterations in the initial conditions of the system affect significantly the evolution, showing that the modeling of the immune system, through non-linear coupled maps, represents a good reproduction of a complex biological system, such as the immune memory. The results for doses at extreme limits, with peculiar behaviors, will be treated in the continuation of this paper.

In Figure 1 the time evolution of the first clonal population that recognizes the first inoculated antigen, is shown: (a) with the addition of antibodies population, and (b) without considering antibodies in the set of coupled maps. The addition of antibodies to the system do not originate a considerable local disturb, however, in Figures 1 to 3, it is shown that the addition of antibodies alters the global capacity of the immune memory, for different antigenic concentrations.

The results obtained without the presence of the term referring to the antibodies. Figure 1(b), correspond to the simplified model (LAGRECA, ALMEIDA, SANTOS, 2001), in which the antibodies soluble in the blood are not considered.

In the Figures 1 and 2 the memory capacity is represented, considering the system with or without the presence of antibodies. For both concentrations of antigens – 0.08 and 0.10 – when the populations of antibodies are considered (Figures 1(a) and 2(a)), the capacity of the immune memory network is smaller than in the absence of populations of antibodies (Figures 1(b) and 2(b)).

In the Figure 4, with high antigenic dosage, it is possible to notice clearly that the bigger the antibody proliferation rate, the smaller the network capacity. In the absence of a specific model for the antibodies, the populations reach higher levels.

Taking into account that the populations of antibodies soluble in the blood help in the regulation of the B cells differentiation, we can infer from the results not only the important role of the antibodies in the mechanism of regulation of the proliferation of the B cells, but also in the maintenance of the immune memory.

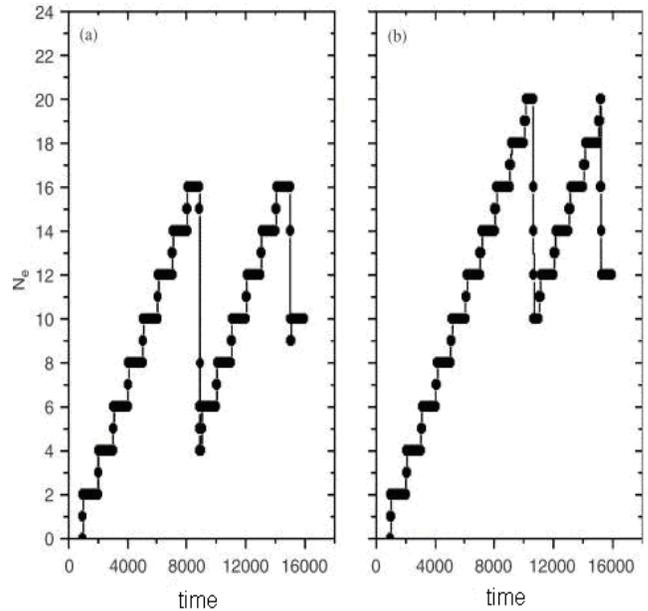

**Figure 2** – Memory capacity for the concentration of antigens equals 0.08: (a) with the addition of antibody population, and (b) without antibodies.

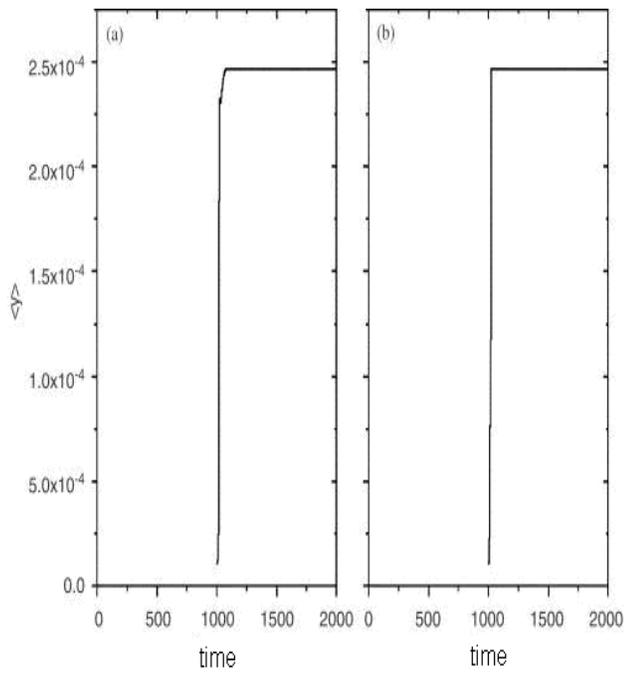

**Figure 1** – Time evolution of the first clonal population that recognizes the first inoculated antigen with (a) addition of antibody population and (b) without antibody.

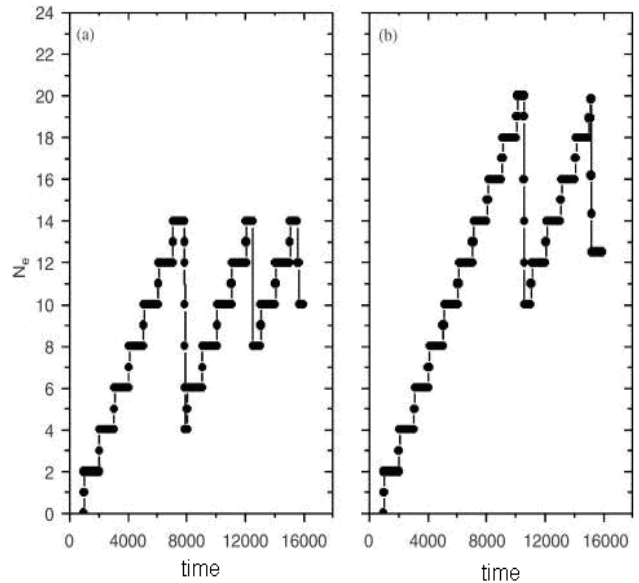

**Figure 3** – Memory capacity for antigenic concentration equals 0.10: (a) with the addition of antibody populations, and (b) without antibodies.

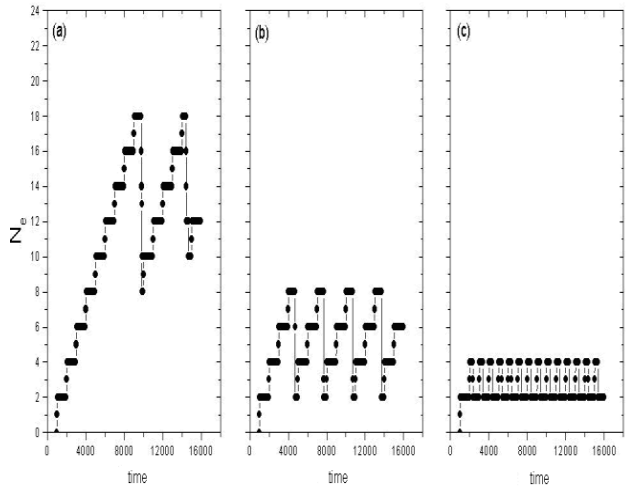

**Figure 4** – Memory capacity for the concentration of antigens equals 1.0: (a) without the addition of antibody populations; (b) with antibody proliferation rate equals 100 and (c) with antibody proliferation rate equals 5000.

The decrease of active populations can be explained through the interaction between antibodies and B cells, which is according to the immune network theory. The results suggest that, despite promoting the fighting to infections, the high production of antibodies can destroy the memory clonal populations produced by previous infections.

Though the dynamical model proposed, we also have reproduced *in maquina* , experiments to study the behavior of the system facing antigenic mutation. Using 10 samples that represent organisms with the same initial conditions, the several populations of antigens are inoculated with concentrations fixed in 0,1and injected in intervals of 1000 time steps. When a new antigen is introduced, its interactions (connections) with all the other components in the system are obtained according to a random number generator. Changing the seed of the random number generator, the bits in the bit-strings are altered (flipped) and, as the bit-strings represent the antigenic variability, the alterations of bits, consequently, represent the respective mutations.

To study the behavior of the system facing mutation, we fixed in 350, 250 and 110 the number of injections of different mutated antigens. The same values of the parameters previously used were considered, which means, the apoptosis or cell's natural death rate $d = 0.99$, clonal proliferation rate equals 2.0 and antibody proliferation rate equals 100. The connectivity parameter $a_h$ was considered equals 0.01 and the bone marrow term $m$ was fixed in $10^{-7}$. Figure 5 shows the average lifespan of the clonal populations that specifically recognized the mutated antigens, considering 350, 250 and 110 injections. The averages, calculated over the 10 samples, indicate that, apparently, the first populations tend to survive more than the others, independently of the number of the inoculations.

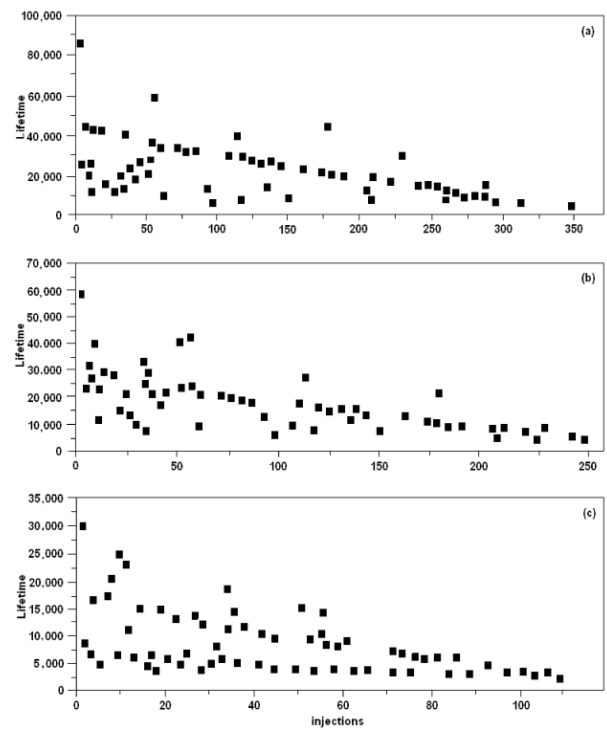

**Figure 5** – Average of the lifespan of the populations that recognized the antigens, for (a) 350, (b) 250 and (c) 110 inoculations.

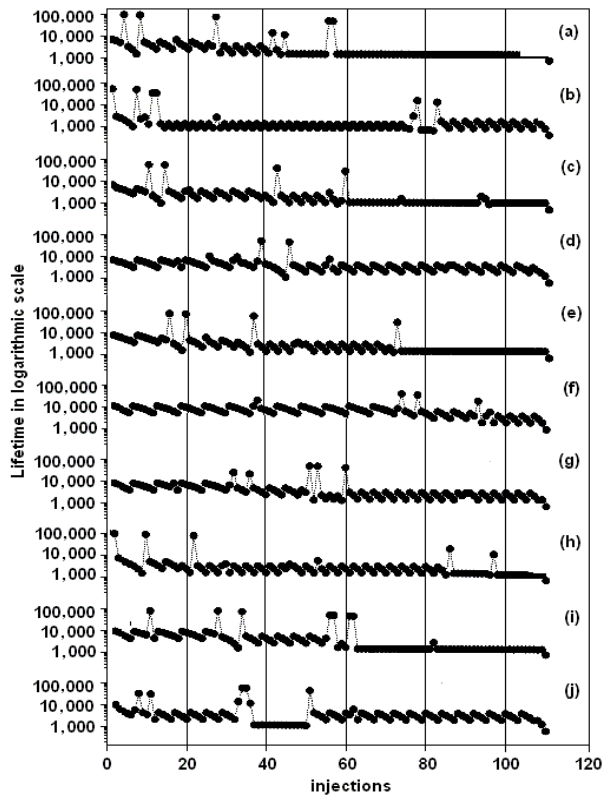

**Figure 6** – Lifespan of the clonal populations in each sample.

However, in Figure 6(a)-(j) it is possible to visualize the behavior of each one of the 10 samples separately, when we administer *in silico* 110 injections. It is clearly noticed, according to Figure 6, that we can not affirm that the first clonal population lasts longer than the populations subsequently recognized other antigens, since the simulations indicate that only in 2 samples the first clonal population have survived for a long time (Figures 6(b) and (h)). It is important to highlight that this discrepancy between the results of Figures 5 and 6 is due to the fact that in two samples the lifespan of the first clonal population excited was long, so, the arithmetic mean was high – even that in other samples the first clonal populations have not survived for a long period.

## 4. DISCUSSION AND CONCLUSIONS

The results presented in this article suggest that the process of antigenic mutation have relation with the durability of the immune memory and the populations of antibodies soluble in the blood not only participate of the immune response, but also help in the regulation of the B cells differentiation. The presence of soluble antibodies change the global properties of the network, and this behavior can only be observed when the populations are treated separately.

The approach proposed also shows that it is impossible to foresee the population that will survive for a long period. These results are reasonable because it is impossible to predict the duration of a vaccine. On the other hand, our results have strongly suggested that the absence or decrease of antibody production promotes the global maintenance of immunizations

## 5. REFERENCES

**AYMERICH F; SERRA M.** (2006). *An Ant Colony Optimization Algorithm for Stacking Sequence Design of Composite Laminates;* CMES, Vol. 13, No. 1, pp. 49-66.

**BURNET, FM.** (1959) *The Clonal Selection Theory of Acquired Immunity.* Cambridge University Press.

**CASTRO, A** (2005). A network model for clonal differentiation and immune memory. Physica A Stat. Mec. Applic., v. 355, pp. 408-426.

**CASTRO, A** (2006). *Antibodies production and the maintenance of the immunological memory.* Eur. Phys. J. Appl. Phys. 33, p. 147–150.

**CASTRO, A.** (2007) *Random behaviors in the process of Immunological memory. Simulation Modelling Practice and Theory 15, p. 831–846.*

**CASTRO, A; FRONZA, C.F; ALVES, D**. (2009) *Viral mutation and its influence in the time evolution of the immunizations.* Eur. Phys. J. Appl. Phys. v. 47, pp. 31401-31406.

**CASTRO, A; FRONZA, C.F; GIACHETTO, PF. ALVES, D. (**2009). *Influence of the antigenic mutations in time evolution of the immune memory: a dynamic modeling.* Lecture Notes in Computer Science, v. 5676, pp. 133-142.

**CASTRO, L.N.** Fundamentals of Natural Computing: Basic Concepts, Algorithms, and Applications. Chapman and Hall/CRC (2006)

**JERNE, NK.** (1974) *Towards a Network Theory of the Immune System. Ann. Immunol.,* v. 125C, p. 373-389.

**KERH T.; LAI, JS. ;GUNARATNAM, D; SAUNDERS, R.** (2008) *Evaluation of Seismic Design Values in the Taiwan Building Code by Using Artificial Neural Network.* CMES, Vol. 26, No. 1, pp. 1-12.

**LACERDA LA.; SILVA JM** (2006). A *Dual BEM Genetic Algorithm Scheme for the Identification of Polarization Curves of Buried Slender Structures.* CMES, Vol. 14, No. 3, pp. 153-160, 2006

**LAGRECA, MC.; ALMEIDA, RMC.; ZORZENON DOS SANTOS, RM.** (2001) *A Dynamical Model for the ImmuneRepertoire.* Physica A, v. 289, p. 191-207.

**LIAN, YS; LIOU, MS** (2005); *Mining of Data from Evolutionary Algorithms for Improving Design Optimization;* CMES, Vol. 8, No. 1, pp. 61-72.

**LUNDEGAARD, C., LUND, O., KESMIR, C., BRUNAK S.. NIELSEN M.** (2007) *Modeling the adaptive immune system: predictions and simulations Bioinformatics.* v. 23, pp.. 3265–3275.

**MONROY, R., SAAB, R., GODÍNEZ, F**. On Modelling an Immune System. Computación y Sistemas Vol. 7 Núm. 4 pp. 249-259 (2004).

**OISHI, A; YOSHIMURA, S.** (2008). *Genetic Approaches to Iteration-free Local Contact Search*, CMES, Vol. 28, No. 2, pp. 127-146.

**OISHI, A.; YOSHIMURA, S;** *A New Local Contact Search Method Using a Multi-Layer Neural Network.* CMES, Vol. 21, No. 2, pp. 93-104, 2007

**PERELSON, AS.; WEISBUCH, G**. (1997) *Immunology for Physicists Rev. of Modern Physics,* v. 69, n.4, p. 1219-1267.

**ROITT, I.; BROSTOFF, J.; MALE, D.** (1998) *Immunology.* 4th Ed. New York: Mosby.

**SINGH, AP.; MANI, V.;GANGULI R.** (2007). *Genetic Programming Metamodel for Rotating Beams,* CMES, Vol. 21, No. 2, pp. 133-148.

**YANG, C. ;TANG D.; HATSUKAMI T.S.; ZHENG J.; WOODARD P.K.** (2007). *In Vivo/Ex Vivo MRI-Based 3D Non-Newtonian FSI Models for Human Atherosclerotic Plaques Compared with Fluid/Wall-Only Models.* CMES, Vol. 19, No. 3, pp. 233-246, 2007

**YOSHIMURA, S.** (2006). *Multi-Agent based Traffic and Environment Simulator -- Theory, Implementation and Practical Application.* CMES, Vol. 11, No. 1, pp. 17-26.